\documentclass[12pt]{article}
\newcommand{\ARAA}{Ann. Rev. in Astron. and Astroph. }
\newcommand{\AandA}{Astron. \& Astroph. }
\newcommand{\ApJ}{Astroph. J. }
\newcommand{\ApJS}{\ApJ Suppl. }

\begin{document}
{\centering
{\Large\bf 
Model for Synchrotron X-rays from Shell Supernova Remnants in 
Nonuniform Interstellar Medium and Nonuniform Magnetic Field
} \\ 
{O. Petruk}\\
{Institute for Applied Problems in Mechanics and Mathematics,\\ 
	   3-b Naukova St., Lviv 79053, Ukraine \\
           {petruk@astro.franko.lviv.ua}}

}
\vspace{1cm}


Possibility to model the high energy synchrotron emission 
(in X- and $\gamma$-rays) from supernova remnants (SNRs) is an 
important task for modern astronomy and astrophysics, because it 
may be responsible for the nonthermal X-rays and TeV $\gamma$-rays 
observed recently from a number of SNRs. 
This emission allows as to look in the processes of particle 
acceleration on SNR shocks and generation of cosmic rays. 
In this paper, a model for the synchrotron emission 
from shell SNR in nonuniform interstellar medium and nonuniform 
magnetic field is presented. 
This model is a generalization of the model of 
Reynolds \& Chevalier developed for a spherical SNR in 
the uniform medium and uniform magnetic field. 
The model will be used for studies on the thermal and nonthermal 
X-ray images and spectra from nonspherical 
SNRs in different interstellar magnetic field configurations. 



\section {Introduction}
	\label{intro}

For the last few years, a considerable interest has grown in 
the investigation of the synchrotron X-ray emission from 
supernova remnants (SNRs),  
caused by electrons accelerated on SNR shocks 
up to the energy $\sim 10^{13-14}\ {\rm eV}$.  
This emission process is believed to be responsible for 
the nonthermal X-ray (SN 1006, Cas A, Tycho, G347.3-0.5, IC443, G266.2-1.2, 
RCW86) and TeV $\gamma$-ray (SN 1006, Cas A, G347.3-0.5) emission 
observed recently from a number of SNRs. 

Model for synchrotron X-rays from shell SNRs 
developed up to now \cite{Rey-Chev-81,Reyn-98} 
works with uniform magnetic field and 
is based on the simple spherical Sedov model \cite{Sedov} 
for SNR hydrodynamics. 
The most essential simplifications of the Sedov model are 
the uniform interstellar medium and spherical SN explosion. 

Observations certify that SNRs are nonspherical objects with 
complicated distributions of surface brightness. The nonuniform 
interstellar medium are the most important factor in modification of 
the structure and evolution of SNRs. 
We present here a model for the synchrotron emission from shell SNR 
in nonuniform medium and nonuniform magnetic field. 
This model is a generalization of the model \cite{Rey-Chev-81,Reyn-98}. 

\section{Model for synchrotron emission from 
nonspherical SNR in nonuniform magnetic field} 
	\label{Sect-model}

\subsection{Hydrodynamics}

The position $R(\theta,\varphi)$ and velocity $D(\theta,\varphi)$ of 
the shock front as well as the distribution of 
gas parameters inside the volume of SNR depend on the ambient density 
distribution $\rho^o(r,\theta,\varphi)$ and anisotropy of the 
supernova explosion 
energy $E_{\rm sn}(\theta,\varphi)$. It is enough to have parameters of gas 
within $R(\theta,\varphi)$ calculated to model the thermal X-ray emission of 
SNR. 
Besides $\rho^o(r,\theta,\varphi)$ and $E_{\rm sn}(\theta,\varphi)$, 
the synchrotron image and spectrum of SNR are also affected by 
the distribution of interstellar magnetic field ${\bf B}_o(r,\theta,\varphi)$ 
which may also be nonuniform. 

It is assumed hereafter that SNR are on the adiabatic phase of evolution. 
The dynamics of the shock front of nonspherical adiabatic SNR and the 
distribution of gas parameters inside whole shocked volume can be 
described with the hydrodynamic method presented in \cite{Hn-Pet-99}. 
In this method, a three-dimensional 
(3-D) object is divided on a number of 1-D sectors 
and the distributions of parameters are found separately for each of them. 
Note, that the velocity of flow is radial in the hydrodynamic 
method \cite{Hn-Pet-99}
and the direction of a sector and electron velocity in it are 
given by the same angles ($\theta,\varphi$) of a spherical coordinate system 
with the origin in the place of explosion. 

We assume also that $\gamma=5/3$, shock compression ratio $\sigma=4$, 
gyrofactor $\eta=1$, $B_{\rm CMB}/B_o\rightarrow0$%
, 
maximum energy attained before the adiabatic stage $E_{\max,0}=0$ 
(see \cite{Reyn-98} for details). 
We do not consider the effect of the cosmic ray pressure 
what can change $\sigma$. 

\subsection{Emission}

The synchrotron volume emissivity of a fluid element $(r,\theta,\varphi)$ 
is \cite{Ginzb-Syrov-65}
\begin{equation}
S_\nu = C(\alpha) KB_\perp^{(\alpha+1)/2}\nu^{-(\alpha-1)/2},
\end{equation}
where $C(\alpha)$ is a constant, $B_\perp$ is the tangential component
of the magnetic field (perpendicular to the electron velocity in the 
fluid element), 
$\nu$ is a frequency, 
$K$ is the normalization of the sharply truncated electron distribution 
\begin{equation}
N(E)dE=\left\{ 
\begin{array}{ll}
KE^{-\alpha}dE, & E\leq E_{\rm max}\\[4pt]
0, 	   & E> E_{\rm max}\\
\end{array}
\right. .
\end{equation}

Let us consider the distribution of the emission parameters 
{\em over the shell} and their evolution {\em downstream}. 

\subsubsection{Spectral index}

The power index $\alpha$ is constant {\em downstream} because electrons lose 
energy proportionally to their energy (Eq.~(\ref{ad-en-loss})) 
and remain essentially confined to the fluid element 
in which they were produced. 
The power index $\alpha$ 
is also constant {\em over the surface} of a nonspherical 
SNR because the shock is strong. Namely, in the first order 
Fermi acceleration mechanism, $\alpha=(\sigma+2)/(\sigma-1)$, where  
the shock compression ratio $\sigma$ does not depend on the ambient density 
distribution in the strong shock limit (Mach number $\gg 1$) realized 
in SNRs: 
$\sigma=\rho_s/\rho^o_s=(\gamma+1)/(\gamma-1)$, 
the subscripts "$s$" and "$o$" refer to the values at the shock and to  
the surrounding medium respectively. 

\subsubsection{Magnetic field components}

The components of the magnetic field evolve differently behind the shock:  
$B_\perp$ rises everywhere {\em at the shock} by 
the same factor $\sigma=\rho_{\rm s}/\rho^o_{\rm s}$ 
and $B_\parallel$ is not modified by the shock 
\cite{Rey-Chev-81}: 
\begin{equation}
B_{\perp,{\rm s}}=B_{\perp,{\rm s}}^o\cdot{\rho_{\rm s}\over\rho^o_{\rm s}}, 
\quad 
B_{\parallel,{\rm s}}=B_{\parallel,{\rm s}}^o. 
\end{equation}
No further turbulent amplification of the magnetic field is assumed.  

{\em Downstream} variation of the components follows from 
the flux-freezing condition
$B_{\perp}(r)rdr=B_{\perp}^o(a)ada$, 
continuity equation 
$\rho(r) r^2dr=\rho^o(a)a^2da$
and magnetic flux conservation 
$r^2B_{\parallel}(r)=a^2B_{\parallel}^o(a)$,
where $a$ is Lagrangian and $r=r(a,t)$ is Eulerian coordinates.  
Thus, if both ${\rm B}_o$ and $\rho^o$ are nonuniform, 
the components of magnetic field ${\rm B}$ are 
\begin{equation}
B_\perp(a,t)=B_{\perp,{\rm s}}(a)\ {\rho(a,t)\over\rho^o(a)}\ {r(a,t)\over a},\ \ 
B_\parallel(a,t)=B_{\parallel,{\rm s}}(a)\left({a\over r(a,t)}\right)^2\!. 
\end{equation}
The components of the ambient magnetic field ${\bf B}_o$ 
in the point ($r,\theta,\varphi$) are 
\begin{equation}
B_\perp^o=B_o\ \!|\sin\Theta_o|,\ \ 
B_\parallel^o=B_o\ \!|\cos\Theta_o|, 
\end{equation}
where the obliquity angle $\Theta_o$ is the angle between 
the directions of the magnetic field ${\bf B}_o$ in this point and 
the electron velocity in the sector
. 

\subsubsection{Normalization $K$}

The assumption that the energy density $\omega$ of relativistic
particles is proportional to the energy density of the magnetic field 
in each fluid element 
\begin{equation}
\omega\equiv\int\limits^{E_{\max}}_{E_{\min}}EN(E)dE=
K\int\limits^{E_{\max}}_{E_{\min}}E^{1-s}dE
\propto B^2
\end{equation}
gives for the variation of $K$ {\em over the shell} and {\em downstream} 
\begin{equation}
K\propto \left\{
\begin{array}{ll}
B^2\big(E_{\max}^{2-\alpha}-E_{\min}^{2-\alpha}\big)^{-1},& \alpha\neq 2 \\[6pt] 
B^2\ln \big(E_{\min}/E_{\max}\big),& \alpha=2.
\end{array}
\right.
\end{equation}

\subsubsection{Injection energy}

Particles start to accelerate on the shock 
when their energy $E$ become more than the injection energy $E_{\min}$. 
The first order Fermi acceleration 
mechanism suggests for $E_{\min}$: 
$E_{\min}=1.68\ \!10^{-4}\ \!\ln\Lambda^{2/3}\ \!
(\lambda_o n_{\rm tot}/D)^{2/3}\ {\rm eV},$ 
where 
$\ln\Lambda$ is Coulomb logarithm, 
$\lambda_o=1.0\ \!10^{18}\ \!T_6^2/n_{\rm tot}\ {\rm cm}$ 
is the mean free path of electron, 
$n_{\rm tot}$ is the total number density (electrons plus ions) \cite{CRbooks}. 
This $E_{\min}$ is $14.65\ln\Lambda^{2/3}$ times the average kinetic 
energy of electrons. 

Variation of $E_{\min}$ {\em over the shell} of nonspherical SNR finds 
for each sector independently: 
\begin{equation}
E_{\min}(R,\theta,\varphi)=
0.026\ \!\ln\Lambda^{2/3}D_3(\theta,\varphi)^{2}\quad {\rm MeV}, 
\end{equation}
where $D_3$ is the shock velocity in $10^3\ {\rm km/s}$. 

If particles move {\em downstream} leaving the region of acceleration, 
the energy varies as \cite{Reyn-98} 
\begin{equation}
E(a)=E_{\rm o}\ \!\overline{\rho}(a)^{1/3},
\label{E-downstr}
\end{equation}
if electron loses its energy due to the adiabatic expansion only 
\begin{equation}
\dot{E}=E\ {\dot{\overline{\rho}}/(3\overline{\rho}}) 
\label{ad-en-loss}
\end{equation}
where $\overline{\rho}(r)=\rho(r)/\rho_{\rm s}$,
$E_{\rm o}$ is the energy of the fluid element $a$ 
at time $t_o$ when the shock crosses it. 

\subsubsection{Maximum energy}

The maximum energy $E_{\rm max}$ which electrons are accelerated to 
varies {\em over the shell} of nonspherical SNR. 
Therefore $E_{\rm max}$ also finds independently in each sector, with 
the expressions given in \cite{Reyn-98}: 
$E_{\max}(R,\theta,\varphi)=\min(E_{\max,1}, E_{\max,2}, E_{\max,3})$ where   
the radiative losses, finite time of acceleration and escaping of 
electrons allow for the maximum energies: 
\begin{equation}
E_{\max,1}=
 7.8\ \! 10^{13}\ \!F_1\ \!B_{o,{\rm \mu}}^{-1/2}\ \!D_3\quad {\rm eV},
\label{Emax1}
\end{equation}
\begin{equation}
E_{\max,2}=
 1.9\ \! 10^{14}\ \!
 \int_{t_{\rm ad}}^{t}\!\! 
 F_2\ \!B_{o,{\rm \mu}}\ \!
 D_3^2\ \!dt_3\quad {\rm eV},
\label{Emax2}
\end{equation}
\begin{equation}
E_{\max,3}=
 7.7\ \!10^{12}\ \!\lambda_{\max,17}\ \! 
 B_{o,{\rm \mu}}\quad {\rm eV}.
\label{Emax3}
\end{equation}
Here the age of SNR $t$ and the time of transition into the adiabatic stage 
$t_{\rm ad}(\theta,\varphi)$ are in $10^3$ years, $B_{o,{\rm \mu}}$ is 
$B_{o}$ in ${\rm \mu G}$, 
$\lambda_{\max,17}$ is $\lambda_{\max}$ in $10^{17}\ {\rm cm}$.  
This $\lambda_{\max}\sim10^{16-18}\ {\rm cm}$ is a free parameter.                  

The geometric factors are \cite{Reyn-98} 
\begin{equation}
F_1=\sqrt{G/R_j},\quad F_2=1/R_j,
\end{equation}
with 
\begin{equation}
 \begin{array}{l}
 \displaystyle 
 G={Z+4/\sigma_B\over Z+4\sigma_B},\quad 
 Z={\cos^2\Theta_{o}+1\over \cos^2\Theta+1}, 
 \\ \\ \displaystyle 
 R_j={1\over 2}{\sigma_B(\cos^2\Theta_{o}+1)+4(\cos^2\Theta+1)\over 
 4+\sigma_B},
 \\ \\ \displaystyle 
 \sigma_B\equiv{B_{\rm s}\over B_{\rm s}^o}=\sqrt{1+16\tan^2\Theta_{o}
 \over 1+\tan^2\Theta_{o}},
 \end{array}
\end{equation}
where the obliquity angles are 
$\Theta_{o}$ for upstream and $\Theta$ for downstream; 
$\tan\Theta=4\tan\Theta_o$. 

The maximum energy changes {\em downstream} in accordance to (\ref{E-downstr}). 

\subsubsection{Obliquity angle} 

Synchrotron emission of the fluid element $(r,\theta,\varphi)$ depend on 
the values $B_o(r,\theta,\phi)$ and $\Theta_o(r,\theta,\phi)$. 
The calculation of $\Theta_o$ depend on the way how the distribution 
${\bf B}_o(r,\theta,\phi)$ is given. 

If we know distributions of 
$B_o(r,\theta,\phi)$, $\theta_H(r,\theta,\phi)$ and 
$\varphi_H(r,\theta,\phi)$, where $(\theta_H,\varphi_H)$ are the spherical 
angles of the vector ${\bf B}_o$ in the point $(r,\theta,\phi)$, then 
\begin{equation}
 \cos\Theta_o=
 \sin\theta\sin\theta_H\cos(\varphi-\varphi_H)+|\cos\theta\cos\theta_H|.
\label{Theta-1}
\end{equation}

There are three independent of a coordinate system angles in our task: 
the inclination angle $\delta$ between the line of sight and the density gradient, 
the aspect angle $\phi$ between the line of sight and the ambient magnetic 
field, and 
the third angle $\xi$ between the magnetic field and the density gradient. 

Let as consider two interesting particular cases. 

i). If direction of the ambient magnetic field is uniform, 
i.e. $\phi(r,\theta,\phi)={\rm const}$,  
then let us accept that the 
origin of the coordinates coincides with the place of explosion, 
the axis $x$ of Cartesian coordinate system 
is oriented opposite to the direction of the line of sight 
(toward observer) and that ${\bf B}_o$ lies in the plane $(xz)$. 
In such a case $\cos\Theta_o$ 
is given by (\ref{Theta-1}) with $\theta_H=\pi/2-\phi$ and 
$\varphi_H=0$. 

ii). If direction of the density gradient is uniform, 
i.e. $\delta(r,\theta,\phi)={\rm const}$, 
then let us accept that the 
origin of the coordinates coincides with the place of explosion, 
axis $z$ of Cartesian coordinate system 
is oriented opposite to the density gradient and the line of sight lies in 
the plane $(xz)$. 
Geometrical consideration yields for $\delta\neq0$ and $\xi\neq0$
\begin{equation}
\begin{array}{c}
 \cos\Theta_o=\sin\theta\sin\xi\cos(\varphi-\varphi_H)+|\cos\theta\cos\xi|,\\[4pt]
 {\rm where\ }\cos\varphi_H=
 \displaystyle{\cos\phi-|\cos\delta\cos\xi|\over\sin\delta\sin\xi}.
\end{array}
\end{equation}
If $\delta=0$ or $\pi$ 
(${\rm grad}\rho^o$ is parallel to the line of sight) but $\xi\neq0$, 
then 
$\theta_H=\pi-\xi$ and $-\xi$ respectively, and additionaly 
$\varphi_H(r,\theta,\varphi)$ should be known. 
If $\xi=0$ or $\pi$ (direction of ${\bf B}_o$ coincides with 
${\rm grad}\rho^o$), then $\Theta_o=\pi-\theta$ or 
$\theta$ respectively.

\section {Conclusions}

The model for the synchrotron emission of nonspherical SNR in nonuniform 
interstellar magnetic field is presented. Being applied to studies of high 
energy emission of SNRs, in particular to X-rays, it allows as to analyse 
the thermal and synchrotron X-ray images and spectra of the objects, 
to make conclusions about the 
SNR itself, supernova explosion, structure and properties of the 
interstellar medium, 
shock wave physics, acceleration processes on the shocks, 
$\gamma$-ray and cosmic ray generation. 



\end{document}